\documentclass{aastex}
\usepackage{spr-astr-addons}
\usepackage{url}
\usepackage{epsfig}

\RequirePackage{color}

\begin{document}

\title{The World Space Observatory Ultraviolet (WSO-UV), as a bridge to future UV astronomy}
\slugcomment{Not to appear in Nonlearned J., 45.}
\shorttitle{WSO-UV as a bridge to future UV astronomy}
\shortauthors{Authors et al.}

\author{B. Shustov\altaffilmark{1}}
\affil{Institute of Astronomy of the RAS, Russia}
\and
\author{A.I. G{\'o}mez de Castro \altaffilmark{2}}
\affil{AEGORA Research Group, Universidad Complutense, Spain}
\and
\author{M. Sachkov\altaffilmark{1}}
\affil{Institute of Astronomy of the Russia}
\and
\author{J.C. Vallejo\altaffilmark{2}}
\affil{AEGORA Research Group, Universidad Complutense, Spain}
\and
\author{P. Marcos-Arenal \altaffilmark{2}}
\affil{AEGORA Research Group, Universidad Complutense, Spain}
\and
\author{E. Kanev\altaffilmark{1}}
\affil{Institute of Astronomy of the Russia}
\and
\author{I. Savanov\altaffilmark{1}}
\affil{Institute of Astronomy of the Russia}
\and
\author{A. Shugarov\altaffilmark{1}}
\affil{Institute of Astronomy of the Russia}\and
\and
\author{S. Sichevskii\altaffilmark{1}}
\affil{Institute of Astronomy of the Russia}

\altaffiltext{1}{Institute of  Astronomy, Russian Academy of Sciences, \\
	Pyatnitskaya 48, 119017 Moscow, Russia}
\altaffiltext{2}{AEGORA Research Group, Fac. CC. Matematicas, Universidad Complutense, Plaza de Ciencias 3, 28040 Madrid, Spain}

\begin{abstract}
The ultraviolet (UV) astronomy is a very demanded branch of space astronomy. Many dozens of short-term UV-experiments 
in space, as well as long-term observatories, have brought a very important knowledge on the physics and
chemistry of the Universe during the last decades.  Unfortunately, no large UV-observatories are planned to be launched 
by most of space agencies in the coming 10 -- 15 years. Conversely, the large UVOIR observatories of the future will appear not earlier than 
in 2030s.  This paper briefly describes the projects that have been proposed by various groups. We conclude that 
the World Space Observatory -- Ultraviolet (WSO-UV) will be the only 2-m class UV telescope with 
capabilities similar to those of the HST for the next decade. The WSO-UV has been described in detail in previous publications, 
and this paper updates the main characteristics of its instruments and the current state of the whole project. 
It also addresses the major science topics that have been included in the core program of
the WSO-UV, making this core program very relevant to the current state of the UV-astronomy. Finally, we also present 
here the ground segment architecture that will implement this program.

\end{abstract}


\section{Introduction}

During almost half a century the astronomers have enjoyed a
continuous access to the ultraviolet (UV) domain, 91.2 -- 320\,nm,
where the resonance transitions of the most abundant atoms and ions
at temperatures between 3\,000 and 300\,000\,K reside.
Because this UV-range is not accessible from ground-based facilities, many short-term UV-experiments in space were carried out and  a number of long-term 
space UV-observatories were put into orbit. They   have brought a very important knowledge on the physics and chemistry of the Universe
during the last decades.

Major achievements of UV-astronomy in this half a century are the following:

\begin{itemize}
\item Direct detection of H$_2$ molecules in space with ``Aerobee-150'' (\cite{Carruthers1970}).
\item Discovery of the hot phase of the interstellar medium with ``Copernicus'' (\cite{Jenkins1974}).
\item Measurements of the D/H ratio with "Copernicus" (\cite{Rogerson1973}).
\item Massive accurate determination of the chemical composition of stars and detailed studies of stellar mass loss phenomena across the HR diagram with 
the IUE mission (1978 -- 1996).
\item Identification of Warm-Hot Intergalactic Medium as the reservoir of missing baryons with HST and FUSE.
\item Giant progress in understanding physics and chemistry of comets, planetary atmospheres and exospheres of exoplanets with HST.
\end{itemize}

All UV observations carried out from space, ranging from 
short-lived rocket and balloon-borned experiments, to small-aperture cameras and spectrographs on astrophysical and planetary missions, 
up to long-lifetimes UV-observatories with powerful instruments on-board,
have shown that UV-instruments on-board satellites bring great scientific benefits to the astronomical community. This makes natural a keen demand on new UV observatories. 

The structure of this paper is as follows. In Section 2 we  briefly review the current state  of the art in UV astronomy,
and discuss the ideas and concepts for future UV-observatories and instruments. 
In Section 3 we present current  status of the World Space Observatory -- Ultraviolet (WSO-UV) mission,
which seems to be the only 2-m class UV telescope to fly in space in the next decade.
Section 4 addresses the key science issues of this mission.
The following section, Section 5, describes how the science program  will be managed, and describes the
current status of the WSO-UV ground segment. The last section, Section 6, is devoted to the concluding remarks.

\section{On UV-instruments for astrophysics for today and tomorrow}

The telescopes for ultraviolet, optical and near-infrared ranges are commonly combined into the so-called UVOIR telescopes. 
This is because their optical design, detectors and general operational conditions  (e.g. cryogenic temрeratures are not required) are fundamentally similar. 
In this section discuss the current and future UVOIR facilities that are fully or partially 
intended for astrophysical observations in the near (170 -- 320\,nm ) and far (91 -- 180\,nm) UV wavelength range.
Regarding the extreme UV, the effectiveness of instruments for  astrophysical observations in 
the range $\lambda<$91\,nm is strongly limited by the high opacity of the interstellar medium, and after the EUVE (Extreme Ultraviolet Explorer) mission (\cite{Bowyer1991})
there is no future significant project for observations beyond the Solar System in the extreme UV.

There is only one large UVOIR observatory currently flying in space, the Hubble Space Telescope (HST). In addition, there are 
three UV instruments on-board astrophysical observatories. The first instrument to consider is 
the Ultraviolet and Optical Telescope (UVOT), on-board the SWIFT satellite. UVOT  has a 30 cm aperture that provides a $17'\times17'$  field of view with a spatial resolution of 0.5'/pixel in the optical/UV band (\cite{Roming2005}).  The UVOT is based on the XMM-Newton mission's Optical Monitor  instrument (\cite{Mason2001}), with improved optics and upgraded onboard processing computers. The third  instrument is the Ultra Violet Imaging Telescope (UVIT) on-board ASTROSAT satellite.  UVIT consists of twin 38-cm telescopes -- one for the FUV region  and the other for the NUV  and visible (VIS) regions. UVIT is primarily an imaging instrument, simultaneously generating images in the FUV,
NUV and VIS channels over a  field of diameter 28 arcmin. Detailed description and review of the first results are presented in \cite{Subramaniam2016}.

It is also of interest to report that the first China's lunar rover, launched in December 2013, is equipped with a 150\,mm Ritchey-Chr\'etien telescope, called LUT (Lunar Ultraviolet Telescope).  This telescope is being used to observe galaxies, active galactic nuclei, 
variable stars, binaries, novae, quasars and blazars in the near UV band (\cite{Wang2015}).
It is capable of detecting objects at a brightness as low as magnitude 13, and the thin exosphere and slow rotation of the Moon allow extremely long, uninterrupted observations of the observed targets. Hence, the LUT can be considered the first long term lunar-based astronomical observatory. 

Some other planetary missions are equipped with UV-instruments, which are in use in Target of Opportunity (ToO) mode. But we do not discuss 
here these missions, neither we will discuss the existing solar physics missions, confining ourselves to astrophysical projects.

As to the future a lot of ideas and projects are under discussion. Recently, \cite{Scowen2017} presented the science cases and technological discussions 
that came from the workshop titled ``Finding the Ultraviolet-Visible Path Forward'', 
held at NASA GSFC on June 25 -- 26, 2015. The material presented there outlined the compelling science that can be enabled by a 
next generation of space-based observatories dedicated to UV-visible science, the technologies that are available for the design 
of those observatories, and the possible launch approaches to enhance the returned UV-science. A number of very large (apertures $>$4\,m) UVOIR space telescopes have been 
proposed and intensively discussed during the last two decades. 
Any future UV/Optical telescope will require these large apertures to complement 
the very large IR and X-ray space telescopes and the 30\,m-class ground-based telescopes that  will arrive in near future. 
Such large projects  were considered at the Kavli IAU Workshop 
on ``Global Coordination: Future Space-Based Ultraviolet-Optical-Infrared Telescopes'', 
held in Leiden, the Netherlands, on July 17 -- 19, 2017. The workshop sessions 
centered on 4 themes: (1) Setting the stage -- astronomy in
the 2030s, (2) Science drivers -- why do we need a large UVOIR mission?, 
(3) Visions for large missions -- what capabilities do we need?, and 
(4) How large missions fit into long-term space mission plans\footnote{The materials and results of the workshop are available at https://www.strw.leidenuniv.nl/KavliIAU2017}.  
  
New technologies required for future large UV telescopes are not still all mature. The problem of very high cost remains unsolved either. It seems that the epoch for operational huge space UV-telescopes will come not earlier that in 2030s. 

Despite of importance of UV observatories for the science progress, there are no consolidated plans in the programs of the major space agencies concerning even for mid-class UV-instruments for the next decade, but the the ``World Space Observatory -- Ultraviolet'' (WSO--UV) project. WSO-UV  
seems to be the only 2\,m-class UV-telescope with capabilities similar to those of the HST
that will guarantee access to the UV wavelength domain during the next decade. 
The WSO-UV is well developed. Therefore, it can be considered as a bridge from present facilities to future UV-astronomy obsevatories.

\section{The World Space Observatory -- Ultraviolet mission}

The WSO-UV is a multi-purpose international 
space mission born as a response to the growing up demand for UV-facilities
by the astronomical community.
The nominal lifetime of this mission is 5 years, with a planned extension to cover 10 years
of operations.
The name ``World Space Observatory -- Ultraviolet'' was introduced by 
\cite{Wamsteker1999}, and it reflects the international nature of the observatory. The project was initially considered by 
ESA and Roscosmos as a possible development 
of two successful UV projects, the IUE mission (\cite{Macchetto1976, Nichols1996}) and the 
``Astron'' mission \cite{Boyarchuk1986,Boyarchuk1994}. 
Nowadays, the WSO-UV implementation is led by the Federal Space Agency (Roscosmos, Russia), with Spain as a project partner. In the Federal Space Program, which is the major planning document for space activity in Russia, 
 the project is also known as ``Spektr-UF'', because the WSO-UV forms part of the  ``Spektr'' series of space missions. Besides the ``Spektr-UF'' the series 
 includes the ``Spektr-R", or ``Radioastron'' mission, launched in 2011,
and the ``Spektr-RG'' mission, planned for launch in 2019.
According to the Federal Space Program the launch of the WSO-UV is scheduled for 2023. We are to reserve up to six months for tests of the S/C systems and in-flight calibrations of scientific instruments. 

The observatory will be put in a circular geosynchronous orbit 
 with 51.6 degrees inclination;
Earth occultation will be small and the orbital period
will allow to time-track targets and to have rapid access to
targets of opportunity. ``Rapid'' means that we can change program to the Director Discretionary Time Program 
(see below in Sect.5)  in two hours.

The WSO-UV was described in detail in previous publications. See, for instance,  
\cite{Shustov2011, Shustov2014, Sachkov2014,  Sachkov2016}. 
Therefore, we will just present here the most relevant new information and the latest updates to  
the project status. We will also present a brief overview of the major science topics 
that have been included in the Core Program of the WSO-UV, focusing on the recent suggestions for this program.

The funding in Russia for the WSO-UV project is guaranteed, and there are no critical technical problems in 
the implementation the project. 
The telescope is being planned for a launch from Baikonur (Kazakhstan) using a ``Proton'' rocket.
The WSO--UV will use the Russian NAVIGATOR bus (platform), designed
by Lavochkin Association. This platform has been already tested successfully in-flight 
by the ``Radioastron'' mission, and it has also been used by some commercial satellites.

\begin{figure}[!ht] 
\begin{center}
\includegraphics[width=0.9\linewidth] {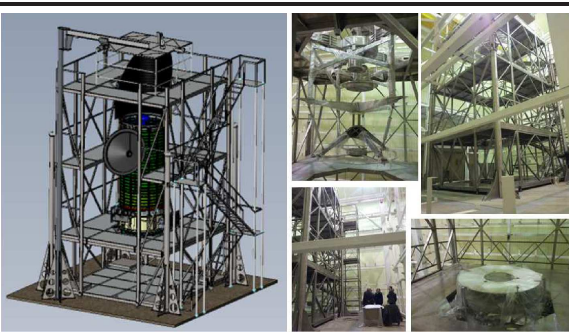}
\caption{The assembly stand of the T-170M telescope.}
\label{stend}
\end{center}
\end{figure}

The WSO-UV consists of a 1.7\,m aperture Ritchey-Chr\'etien T-170M telescope  (focal length  17\,m, field of view 30 arcmin).
The mechanical part of this telescope has been manufactured by Lavockin Association, and it has
successfully passed the required dynamic and thermal tests. The optics of the telescope has been manufactured by the 
Lytkarino optical plant, and both optics and coatings (Al+MgF$_2$) meet the technical requirements. 
The AIV (Assembly Integration and Verification) facilities are almost ready to use, and the pre-final state of one of these AIV  elements, an assembly stand for the T-170M 
telescope, is shown in Fig.~\ref{stend}.

The telescope is  equipped with instrumentation designed to carry out high resolution spectroscopy, 
long-slit low resolution spectroscopy and direct sky imaging.
The main science instruments are the WSO-UV spectrographs (WUVS) and the Field Camera Unit (FCU).

The optical design of WUVS is presented in \cite{Panchuk2014}. The WUVS consists of the following three channels (spectrographs):
\begin {itemize}
\item The FUV high resolution spectrograph (VUVES), to carry out echell\'e spectroscopy with high resolution ($R\sim 50\,000$) in the 115 -- 170\,nm range.
\item The NUV high resolution spectrograph (UVES) to carry out echell\'e spectroscopy with $R\sim 50\,000$ in the 174 -- 310\,nm range.
\item The Long Slit Spectrograph (LSS) that will provide low resolution ($R\sim1000$), long slit spectroscopy in the 115 -- 305\,nm range. The spatial resolution will be better than 0.5\,arcsec (0.1\,arcsec as the best value).
\end{itemize}

\begin{figure}[!ht] 
\begin{center}
\includegraphics[width=0.9\linewidth] {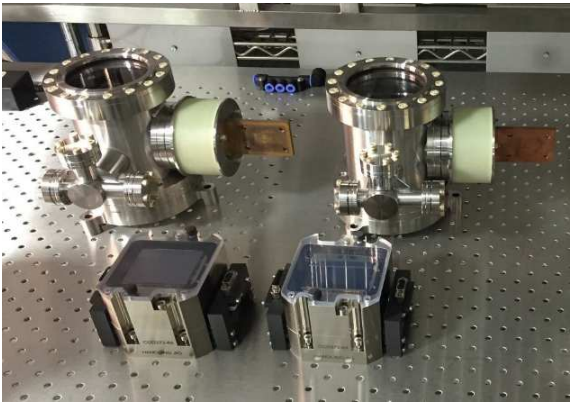}
\caption{Qualification Engineering Models of CCD detectors.}
\label{ccd}
\end{center}
\end{figure}

All spectrographs will be equipped with CCD detectors \cite{Shugarov2014}. The Fig.~\ref{ccd} shows the 
Qualification Engineering  Model of these CCD detectors.

The spectral resolution provided by VUVES and UVES channels is similar to that
offered by STIS/HST, but higher than the maximum resolution provided by COS/HST ($R\sim$20\,000).

\begin{figure}[!ht] 
	\begin{center}
		\includegraphics[width=1.0\linewidth] {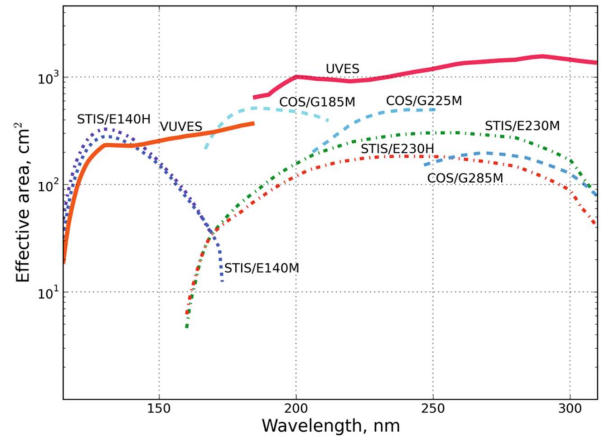}
		\caption{The estimated effective area of UVES and VUVES channels of the WUVS, compared with the COS/HST and STIS/HST channels.}
		\label{WUVS_comparision}
	\end{center}
\end{figure}

\begin{figure}[!ht] 
	\begin{center}
		\includegraphics[width=1.0\linewidth] {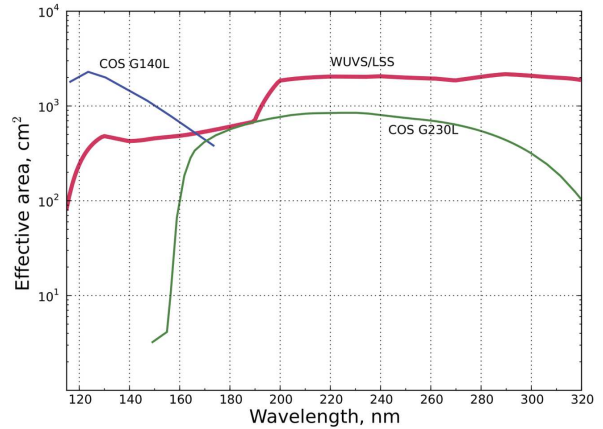}
		\caption{The estimated effective area of the LSS channel of the WUVS, compared with the COS/HST channels.}
		\label{lss}
	\end{center}
\end{figure}

The estimated effective area of UVES and VUVES channels of the WUVS
is presented in Fig.~\ref{WUVS_comparision}, along the same parameters for the COS/HST and STIS/HST channels, for comparison purposes. 
Similarly, the estimated effective areas of the LSS channel of the WUVS, 
and the corresponding channels of the COS/HST are presented in Fig.~\ref{lss}.

The imaging functions of the WSO-UV were initially assigned to the Imaging and Slitless Spectroscopy Instrument (ISSIS),
under responsibility of Spain \cite{GomezdeCastro2014}, 
designed to carry out imaging and slitless spectroscopy in the 115 -- 320\,nm spectral range. 
The ISISS was to be equipped with two MCP detectors, with CsI and CsTe photocathods for FUV and NUV observations, 
 respectively. The capabilities of the ISSIS was expected to be similar to those of the Advanced Camera for Surveys of the HST (ACS/HST). 

\begin{figure}[!ht] 
\begin{center}
\includegraphics[width=0.9\linewidth] {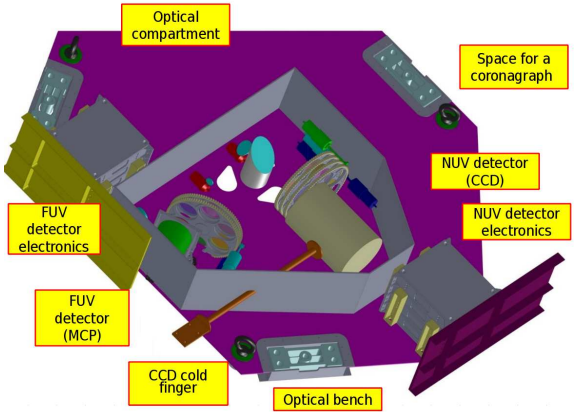}
\caption{Preliminary design of the FCU.}
\label{FCU}
\end{center}
\end{figure}

This scenario has changed, and now the imaging capabilities for the WSO-UV are fulfilled by the Field Camera Unit (FCU), the successor to ISSIS. 
The FCU will be the first UV-imager to be flown to a geosynchronous orbit above the Earth geocorona. 
The main task of the FCU is to obtain high resolution images and low resolution spectra in FUV and NUV bands \cite{Sachkov2017}. 
The Fig.~\ref{FCU} shows the preliminary design of FCU.

The FUV channel is equipped with a MCP detector sensitive in 115 -- 176\,nm range. The main characteristics of this channel are:

\begin {itemize}
\item Diffraction-limited image quality in FUV.
\item High sensitivity photon-counting mode.
\item High time resolution.
\end{itemize}

The NUV channel is equipped with a CCD detector sensitive in 174 -- 310 (extended to 1000)\,nm range. The main characteristics of this channel are:

\begin {itemize}
\item Wide field of view.
\item High dynamic range. 
\item Possibility of low resolution field spectroscopy (provided by grism).
\end{itemize}

The Table~\ref{fcutable} shows the main features of the FCU channels. This table also presents
the corresponding features of HST/ACS/SBC and HST/WFC3/UVIS, taken from the HST Instruments Handbook, for comparison purposes.

\begin{table*}
\small
\caption{Main features of the WSO-UV FCU, compared with those from HST/ACS/SBC and HST/WFC3/UVIS}
\label{fcutable}
\begin{tabular}{@{}lrrrrrrrrrrr@{}}
\tableline
Parameters & FUV Channel & NUV Channel & HST/ACS/SBC & HST/WFC3/UVIS  \\
\tableline
Detector & MCP & CCD & MCP,	MAMA & CCD\\
Spectral range, nm & 115\sbond 170 & 174\sbond 310 (ext. to 1000) & 115\sbond 170 & 200\sbond 1000 \\
Effective area,	m$^2$ & 0.068 &0.27 &0.18 &0.45\\
Field	of	view, arcsec & $121\times 121$ &$597\times 451$	  &$35\times 31$  &$162\times 162$ \\
Scale, arcsec/pixel  & 0.08$^{[a]}$ & 0.146 &$0.033\times 0.030$	 & 0.0395 \\
Detector	size,	mm & 30	 & 49 $\times$ 37 &25	 & 61 $\times$ 61 \\
Detector format & 2k $\times$ 2k & 4k $\times$ 3k	 &1k $\times$ 1k & 1k $\times$ 1k \\
Number	of	filters & up	to 10 +2 prisms & up to 15 &6+2	prism & 62 \\
\tableline
	[a]  Scale is tentative. & & & & \\
\end{tabular}
\end{table*}

\section{WSO-UV Key Science Issues}
The present day hot science topics should constitute the core program of any space project, and WSO-UV is not an exception. 
It is important to focus the science program of any mission on the most challenging science issues that can be most efficiently studied with the observatory instruments. 
As these science issues evolve with time, any project needs to continuously adapt the mission priorities to the most actual topics. 
The science case of the UV astronomy, and that of WSO-UV in particular, has been 
 discussed recently in \cite{Boyarchuk2016}.
 
Currently, the Core Program of the WSO-UV includes:

\begin{itemize}
\item The determination of the diffuse baryonic content in the Universe and its chemical
evolution. The main topics will be the investigation of baryonic content in Warm-Hot Inter Galactic Matter (WHIM) and the Ly-$\alpha$ systems.

\item The study of the formation and evolution of galaxies.

\item The physics of accretion and outflows: stars, black holes, and all those objects dominated by accretion mechanisms.
 The efficiency and time scales of the phenomena will be studied, together with the role of the radiation pressure and 
 the disk instabilities.

\item The investigation of the extrasolar planetary atmospheres and astrochemistry in
presence of strong UV radiation fields.

\end{itemize}

\subsection {The baryonic content of the IGM}
As it was noted by \cite{Nicastro2014}, baryons are missing at all astronomical scales in  Universe.  Independent of the different models for the 
 evolution of the Universe, the major baryonic component of the Universe  must be associated with the intergalactic medium (IGM). Hydro-dynamical 
simulations for the formation of structures tend to re-concile the different ``missing-baryon'' problems and predict 
that most of the baryonic matter of the Universe is hiding in a hot and tenuous gaseous phase.
 \cite{Bregman2015} argues that about 30--50\% of the baryons in the local Universe  are likely in a hot phase and Warm-Hot Inter Galactic Matter (WHIM),
$T = 10^5 - 10^8$\,K,  and the dominant ions have their ground state resonance transitions in the UV  and X-ray domains.  

Share of missing baryons in the local Universe ($z\le2$) is  almost two times larger than that in the more distant Universe (\cite{Nicastro2014}). 
This is because the distant objects, which emit/absorb in UV, are observable in optical range due to cosmological shift. 
 
Thus UV emiting/absorbing plasma (WHIM and Ly$\alpha$-absorbers) is a major reservoir of missing baryons. 
Observation of broad hydrogen Ly$\alpha$ absorption is one of the most promising techniques to map the distribution of the WHIM.

\subsection {The baryons in DM halo and formation of galaxies.}
The models of the formation of galaxies like our Milky Way
have  predicted that the central galaxy only contains
a modest fraction of the available baryons (\cite{Klypin2011}). Galaxies are inefficient
producers that have converted a small portion of their
available gas into stars. It has become increasingly
apparent over the past years that galaxies also exhibit a diffuse baryonic
component within the dark matter halo that extends far
from the inner regions to the virial radius and beyond. This gas is a target of COS-Halos Project (HST).
\cite{Werk2014} analyzed the physical conditions of the cool, photoionized ($T\sim10^4$K) circumgalactic medium
(CGM) using the COS-Halos suite of gas column density measurements for 44 gaseous halos within
160 kpc  galaxies at $z\sim0.2$, and the WSO-UV is believed to be an efficient instrument to contribute 
to the inventory of baryons in the galactic vicinity.

\subsection{Astronomical engines}
The astronomical engines (stars, black holes, etc...) can accelerate
large masses to velocities close to the speed of light  or
generate sudden ejections of mass as those observed in Supernova
explosions. The WSO-UV will provide key inputs to answer the fundamental open questions
concerning the physics of these objects. Some few examples are:

\begin{enumerate}
\item High resolution UV spectroscopy will allow to determine
the structure of the accretion flow on
magnetic cataclismic variables  and on T Tauri stars, and to
measure the physical conditions and clumpyness of the outflows.
It will also allow to study the source of the energy
that powers the extended dense ($\geq 10^{10}$cm$^{-3}$)
and hot (T$_e \geq 60,000$~K) envelopes 
with luminosities about 0.2 L$_{\odot}$
that have been detected  around T Tauri stars (\cite{AIG2003}).

\item Low resolution spectroscopy will allow to measure
the general physical conditions and
metalicities of the Broad
Lines Emission Region in Active Galactic Nuclei. Reverberation
mapping will allow to study the kinematics and the mass of the
central supermassive black holes. The atmospheres
of the hot accretion disks in cataclysmic variable stars
and the role of disk instabilities in triggering the outbursts
will be also studied.

\item The high sensitivity of the FUV camera will allow to
detect hot jets through their Ly-$\alpha$ and
to resolve the thermal structure of the jets and
the regions shocked by them. It will also allow to
survey star forming regions to detect planetary-mass
objects in regions like $\sigma$-Orionis and to study
the magnetic activity and accretion processes in
these free-floating planetary-like objects which
are at the low mass end of the molecular clouds
fragmentation scales.

\end{enumerate}

\subsection{Extrasolar planetary atmospheres}
The instrumentation of the WSO-UV project is very important and helpful for exoplanet studies and 
for the characterization of the exoplanet-stellar environment.

There are still several difficulties that are at the origin of the major 
uncertainties on any estimation of exoplanetary atmosphere properties (\cite{Fossati2014}):

\begin{itemize}
\item The relative faintness of the UV stellar emissions.
\item The variability of the sources.
\item The signal contamination by both the sky background signal (at some spectral lines) and the instrument
response.
\end{itemize}

While the first difficulty can be solved by focusing on a few close-by
and UV-bright stars, the signal variability from both the source
and the instrument is a real problem that should be addressed to
build a reliable diagnostic to extract an accurate description of
the upper atmosphere and of the interaction region between the
exoplanet and the impinging wind from its host star.

In addition to standard exoplanet observations and characterizations, the WSO-UV
observatory can be also used for observations of biomarkers. The biomarkers like ozone have very
strong transitions in the ultraviolet. These are electronic molecular transitions, hence several
orders of magnitude stronger than the vibrational or rotational transitions observed in the infrared
or radio range. The spectral resolving power required to detect biomarkers in the atmosphere of exoplanets is not
a crucial capability. A resolution of $R\sim\,10000$ is adequate
for these investigations, and even $R\le\,1000$ could be enough to detect the broad band signatures of
many molecules. The presence of biomarkers and other constituents in the atmospheres can be searched by the WSO-UV high
resolution spectrographs for hundreds of  exoplanets orbiting K, G and F-type main sequence stars\index{K stars}\index{G stars}\index{F stars}.

\subsection{Physics of  stars as progenitors of some GW events}
\cite{Hamann2017} noted that the recent discovery of a gravitational wave from the merging of two black holes, of about
30 solar masses each one, challenges our incomplete understanding of massive stars and their evolution. The UV-observations 
are critical for studying the physics of these massive stars.
Critical ingredients comprise mass-loss, rotation, magnetic fields, internal mixing, and mass transfer in close binary systems. The imperfect
knowledge of these factors implies large uncertainties when modeling the stellar populations and their feedback. 
\cite{Wang2017} describes the UV-behaviour  of GW 170817 and concludes that the soft X-ray/UV emission 
could provide the earliest localization of the corresponding electromagnetic counterparts of 
the gravitational waves from double neutron star mergers.

 have rapid access to
targets of opportunity.

\section{Management of the science program and the architecture of the ground segment of the WSO-UV mission}
The aim of the WSO-UV project is the creation of an international observatory working in the UV range which is inaccesible 
for Earth-based instruments.
The WSO-UV as international observatory will make regular Announcements of Opportunity to the scientific community 
for observing time to be solicited by international teams.  The WSO-UV is expected to carry over three main scientific research programs:
\begin{itemize}
\item The Core Program (CP) that includes the key scientific projects driving the development of the WSO-UV mission.
\item The National Programs (NP) that grant a fraction of the observing time to the Countries involved in the WSO-UV development.
\item The Open Program (OP) to the world-wide scientific community.
\end{itemize}

Additional programs are:
\begin{itemize}
\item The Director Discretionary Time Program, that represents a small fraction of the observatory time and will be managed by the director of 
the WSO-UV observatory to allow a rapid response to unexpected important astronomical events or for other scientific purposes.
\item The Calibration Program, that includes specific observations and measurements to calibrate and ensure the optimal performance of the WSO-UV.
\item The Guaranteed Time Program for the Instrument Teams.
\end{itemize}

Only excellent proposals will be eligible for the``Core Program'' (CP) but there are no limitations in the amount of time or instrumentation.
The proposals to all the core program and remaining observing programs will be submitted to the Science Operations Center (SOC) for evaluation 
of their scientific excellence and technical feasibility. Indeed, it is foreseen that there will be two initial
calls devoted to just implement the core program of the mission, well before the launch date. 
The first CP call will be devoted to those proposals requiring 
observing time with other facilities (preparatory observations), and the second CP call will be devoted to those
proposals not requiring any preparatory observation, and to those that participated in the first call.

Therefore, the elements of the ground segment in charge of handling these early calls must be already developed
and implemented in time, well ahead the launch date. Because of this staggered schedule, an incremental and iterative strategy
for implementing the necessary parts of the ground segment has been set in place.

The WSO-UV ground segment is composed of all the infrastructure and facilities involved in the preparation
and execution of the WSO-UV mission operations. These facilities include all equipment and services needed for the mission 
preparation and to perform the actual  operations: Ground Stations, Mission Control Centre, Science Control Centre, Instrument Teams, etc.

Functionally, the ground segment comprises downlink and uplink systems. The downlink part encompass the systems in charge of the 
monitoring and control of the spacecraft, telescope and instrument, as well as reception, processing
and storage of the scientific data. The uplink chain includes those systems in charge of the reception of proposals for observing 
from the users of the telescope, the systems in charge of planning the sequences of observations and the systems in charge of final conversion into 
telecomands to be uploaded to the spacecraft.

This high-level architecture has been refined as the project has evolved. Previous reports on its progress can be found 
in \cite{Malkov2011, AIG2013}.
This architecture splits the ground segment in two ``standard'' elements, 
a Mission Operations Centre (MOC), and a Science Operations Center (SOC). However, 
some simplifications have been recently done to the existing baseline, aiming to reduce the costs and increase
the scientific return of the mission. 

The Mission Operations Centre (MOC) includes all technical systems, operational procedures and manpower 
related to the provision of the spacecraft status analyses,
spacecraft functioning control, monitoring of the platform, overall payload and ground segment. This includes commanding capabilities for the 
platform, payload and ground segment, control procedure generation and execution,
performance evaluation, low-level mission planning sequences generation,
on-board software management and maintenance and operational database management systems. 

The Science Operations Centre (SOC) includes all technical systems, operational procedures and manpower available required for supporting 
the planning, monitoring and control, and performance evaluation of the payload elements on board the space segment.
The SOC activities cover the science operation planning activities, such as proposal reception and handling, and 
creation of a timeline of observations to be uplinked to the satellite (uplink related responsibilities), 
and it also covers the science data operations, such as science data processing and archiving (downlink related responsibilities).

The initial baseline for the ground segment was based on two complete Missions Operations Centres (MOC), and two complete Science Operations Centers (SOC), 
both sharing the operations on a regular basis. A prototype of this integrated dual ground segment was developed and tested during 2012 -- 2013. 
As a consequence of the test campaign, a re-engineering process took place, aiming to simplify the operations, maximize the available resources and reuse 
the existing multi-mission MOC facilities at Russian side. 

The WSO-UV ground segment baseline now relies in one single MOC, located 
in Moscow (Lavochkin Science and Technology Association). This single entity will be the unique interface between the ground segment and the 
spacecraft in which concerns the low-level telecommanding capabilities, fully under Russian responsibility.

The concept of sharing operations remains for the science-side of the WSO-UV operations, under SOC responsibility.
The sharing of science operations is kept by building a modular SOC, where some elements are unique, meanwhile others can be duplicated.
This allows the reuse of the prototyped architecture and components, but also the removal of a heavy common MOC-SOC infrastructure.

\begin{figure}[t]
\includegraphics[width=1.0\linewidth] {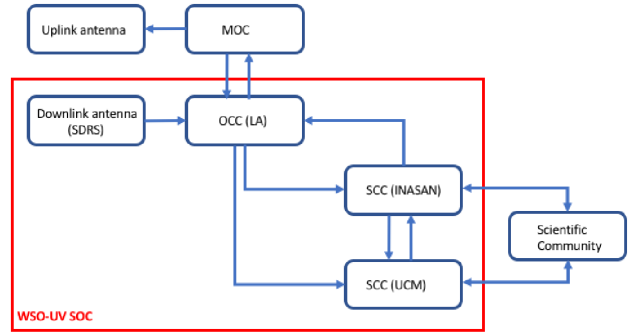} 
\caption{ WSO-UV SOC context within the mission ground segment. LA=Lavochkin. INASAN=Institute of Astronomy of the Russian Academy of Science. UCM=University Complutense of Madrid} 
\label{groundsegment}
\end{figure}

The SOC is responsible 
for the quality of the science mission products and for the scientific performance of the mission. Therefore, its functionalities include
the interaction with the Mission Operations Centre (MOC), instrument calibration, long term planning (integrated science schedule), 
science data archiving, and interaction with the mission external users.
The main SOC elements within the Ground Segment are seen in Figure~\ref{groundsegment}. 
The SOC combines at the same time an uplink function and a downlink function. Therefore, it is split in a Scientific Data Reception System (SDRS),
an Operational Control Centre (OCC), and two Science Control Centres (SCCs).

The SDRS, referred in Russian as ``Apparatus for Reception of Scientific Data'', is formed by the set of mission 
available antennas dedicated to reception of the signal from satellite and sending  the downlink TM data 
(in L0 format) to the MOC and to the Operational SOC.

\bf {ANA's EXPLANATION FOR L1, L1, L2}
\rm

The OCC, referred in Russian as ``Center of Scientific Data Reduction'', is the part of the SOC co-located with
the MOC, and it is formed by the systems in charge of creating the low-level sets 
of sequences that will be converted into TC and uplinked to the S/C by the MOC, receiving the TM data from the 
SDRS and performing a preliminary data reduction (Quick Look Analysis) and sending the downlink TM and Aux data needed 
for scientific processing to the SCC.

The SCCs, referred in Russian as ''Russian/Spanish Regional SOC'', are formed by the set of SOC components 
in charge of the final archive of data, the L2 processing from L1 level, 
the provision of the scientific analysis tools to the community, the provision of proposal reception and handling systems.

There will be two SCCs, one at the Institute of Astronomy of the Russian Academy of Science (INASAN), in Moscow, Russia, and one at the Joint Center for Ultraviolet Astronomy (JCUVA),
located in University Complutense of Madrid, Spain.

Every SCC has the following components:

\begin{itemize}
\item The Mission Data Archive System (MDA), main data repository and index to the WSO-UV mission.
\item The Science Data Pipeline System (SDP), in charge of processing L1 data product as received from dedicated Downlink Antennas 
to obtain mission L2 products. 
\item The Science Control System (SCS), in charge of handling the communications and data interchange 
between the SCC INASAN and the SCC UCM., and the reception of the TM and Aux Data coming from the OCC.
\item The Sequence Generator System (SGS, only at INASAN), that converts the lists of targets approved by the Time Allocation Committee
of the project into low-level sequences of time-tagged set of observations that will be sent to the OCC for later creation of TCs to be uplinked.
\item A set of Interfaces and Scientific Analysis Tools (the IAS), formed by SCC Simulators, Proposal Tools, MDA Access tools and general
Community Support Tools.
\end{itemize}

The SOC conforms one of the major parts of the Spanish contributions to the mission. It is been built 
in a modular and incremental way in order to fullfil the requirements of the mission as they are needed. The 
SCC will be the first operational element of the Ground Segment to be implemented, because it will support the core program call
 to take place well before the launch date of the spacecraft.

The different components of the SCC are defined and developed by a international Science Team 
composed of the Spanish and Russian Science Support Teams based at the Universidad Complutense of Madrid (UCM) 
and the Institute of Astronomy of the Russian Academy of Science (INASAN).
This Science Team provides all technical systems and manpower available required for supporting 
the operation and processing of the Scientific Instruments on-board the WSO-UV spacecraft. Therefore,
it is responsible for  laying the foundation of and supervising all the operations related to the mission primary users: the scientists.
At mission level, the Science Team constitutes the core of the future WSO-UV observatory.

\section{Concluding remarks}
The UV-astronomy is a very demanded branch of space astronomy.   Many dozens of short-term UV-experiments in space as well
 as long-term  space observatories  have brought  very important knowledge on physics and chemistry of the Universe. 
 A number of very large (aperture $>4$\,m) UVOIR space telescopes have been intensively proposed 
 and discussed during the last two decades. Unfortunately, none of these large UV-observatories is planned to be  
 launched in the coming 10 -- 15 years by most of space agencies. Therefore, large UVOIR observatories of the future will not appear earlier than in 2030s. 
The World Space Observatory -- Ultraviolet (WSO-UV) will be the only 2\,m class UV-telescope with capabilities similar to the 
HST ready to fly for the next decade. The WSO-UV will work as a space targeted observatory with a
core program and an open program for scientific projects from
the world-wide community and national (funding) bodies (\cite{Malkov2011}).
The core program of the WSO-UV remains very relevant to the current state of the UV-astronomy.
The latest information on this project can be found at the official web site http://wso-uv.org.

\acknowledgments
B. Shustov and E. Kanev thank the Russian Science Foundation for supporting this work by grant RSF 17-12-01441.
Ana I. G\'{o}mez de Castro,  Juan C. Vallejo and Pablo Marcos-Arenal thank the Spanish Ministry of Economy, Industry
and Competitiveness for grants ESP2014-54243-R and ESP2015-68908-R.

\nocite{*}
\bibliographystyle{spr-mp-nameyear-cnd}
\bibliography{WSO-UV}

\end{document}